\documentclass[12pt]{article}        

\usepackage{epsf}
\usepackage{epsfig}

\usepackage{times}

\usepackage{amsmath}
\usepackage{amssymb}

\usepackage{rotating}

\newcommand{\KK}{${\cal KK}$}
\def\Order#1{${\cal O}(#1$)}
\def\Oceex#1{${\cal O}(#1)_{_{\rm CEEX}}$}
\def\Orderd#1{{\cal O}(#1)}
\def\Oceexd#1{{\cal O}(#1)_{_{\rm CEEX}}}

\begin{document}                     

\begin{titlepage}

\begin{flushright}
{\bf  CERN-TH/99-217 } \\
{\bf  LAPP--EXP--99.05} 
\end{flushright}

\vspace{10mm}

\begin{center}
{\bf\LARGE
  Initial--Final-State Interference \\
  in the Z line-shape }
\end{center}

\vspace{1mm} 

\begin{center}
{\bf S.~Jadach }\\
{\small DESY, Theory Group, D-22603 Hamburg, Notkestrasse 85, Germany}\\
{\small and Institute of Nuclear Physics, ul. Kawiory 26a, 
PL 30--059 Cracow, Poland} \\
{\bf B.~Pietrzyk  }\\
{\small Laboratoire de Physique des Particules LAPP, IN2P3--CNRS,} \\
{\small F--74941 Annecy-le-Vieux Cedex, France}\\
{\bf E.~Tournefier  }\\
{\small CERN, EP Division, CH--1211 Geneva 23, Switzerland}  \\
{\bf B.F.L. Ward  }\\
{\small Department of Physics and Astronomy,}\\
{\small  The University of Tennessee, Knoxville, TN 37996-1200, USA}\\
   {\small and SLAC, Stanford University, Stanford, CA 94309, USA}\\
{\em and}\\
{\bf Z. W\c{a}s  }\\
{\small CERN, Theory Division, CH 1211 Geneva 23, Switzerland,}\\ 
{\small and Institute of Nuclear Physics,
Krak\'ow, ul. Kawiory 26a, Poland}
\end{center}

\vspace{2mm} 
\begin{abstract}
The uncertainty in the determination of the
Z line-shape parameters coming from the precision of the calculation of 
the Initial-State Radiation and Initial--Final-State Interference is 
2$\times$10$^{-4}$ for the total cross section $\sigma^0_{had}$ at the Z peak,
0.15 MeV for the Z mass $M_Z$,
and 0.1 MeV for the Z width $\Gamma_Z$.
Corrections to Initial--Final-State Interference beyond \Order{\alpha^1} are discussed.
\end{abstract}

\begin{center}
{\it Submitted to Physics Letters}
\end{center}

\vspace{1mm}
\begin{flushleft}
{\bf CERN-TH/99-217 \\ July 1999}
\end{flushleft}

\end{titlepage}


High-precision and high-statistics LEP measurements performed at c.m.s.
energies close to the Z mass provide the 
most stringent tests of the Standard Model. In the near future 
LEP experiments will publish results of the final analysis of
the Z line shape. The expected precision of the combined results of the four LEP 
experiments is about 2 MeV on the Z mass $M_Z$ and width $\Gamma_Z$, and 0.1\%
on the $\sigma^0_{had}$ measurement \cite{Karlen}. The 
uncertainty in $\sigma^0_{had}$ includes
an improved precision of the theoretical calculations of the small-angle Bhabha
scattering of 0.061\% \cite{Ward:1998ht,Arbuzov:1996eq}.

Large theoretical corrections are needed to extract the line-shape parameters
$M_Z$, $\Gamma_Z$ and $\sigma^0_{had}$ from the experimentally 
measured cross-sections.
The largest correction, about 30\% on the Z peak, comes from Initial-State 
Radiation (ISR), and it was discussed in Ref.~\cite{StBoMa}.
In view of the high precision of the ISR calculation \cite{StBoMa} 
and of the measurements, the precision
of the calculation of Initial--Final-State Interference (IFI) 
becomes important as well.
Results of independent calculations of this contribution in  
${\cal O}$($\alpha^1$) are compared here.
The new results at higher orders are given.

Two line-shape fitting programs are used by LEP experiments: 
ZFITTER \cite{ZFITTER}  and MIZA \cite{MIZA}.
ZFITTER was cross-checked with TOPAZ0 \cite{TOPAZ0}.
The influence of the precision 
of the IFI calculation on the precision of the fitted Z line-shape
parameters is studied here in the c.m.s. energy region between 88 and 94 GeV.
This study was done with the fitting program MIZA, as used by the ALEPH 
Collaboration. 

\begin{figure}[htb]
  \begin{center}\mbox{
      \epsfig{file=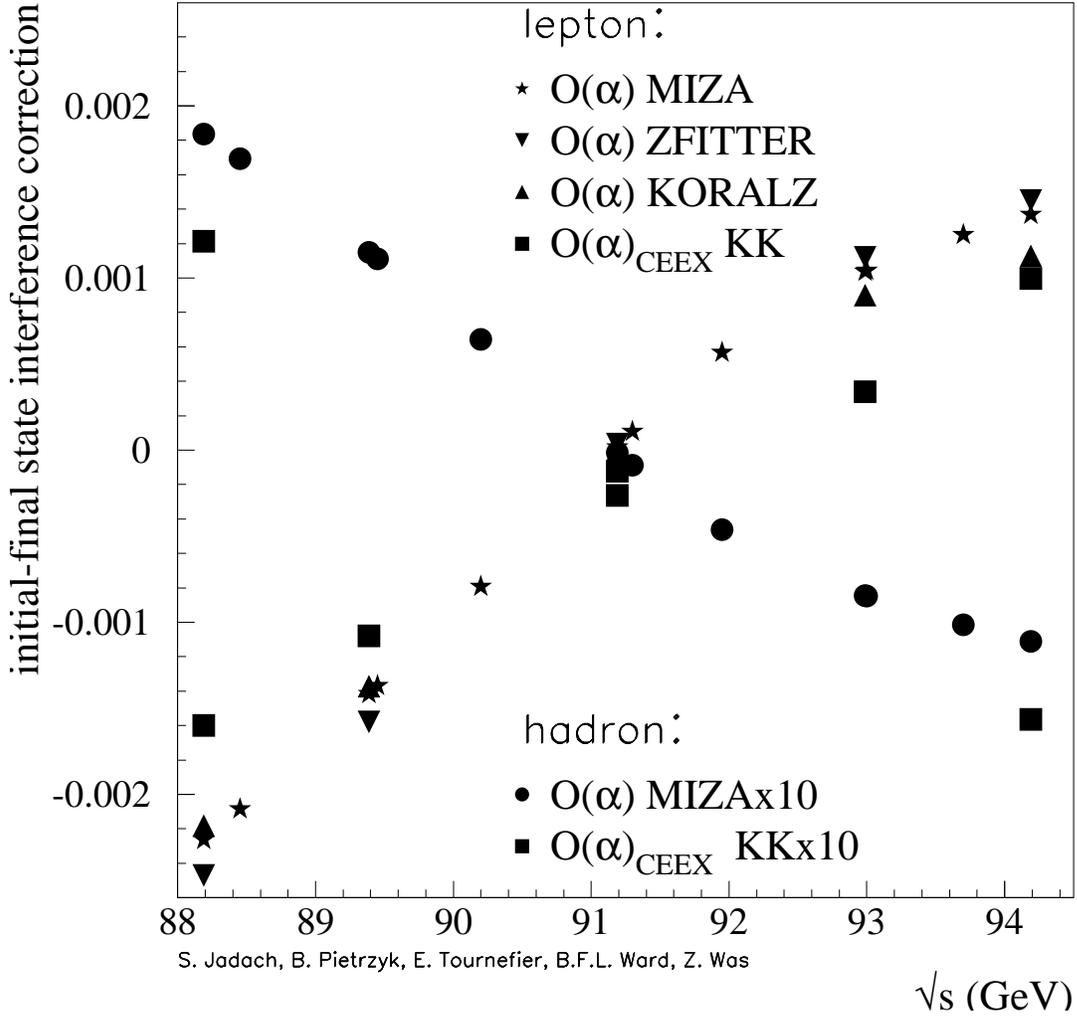,width=15cm}}
  \end{center}
  \caption[]{\sf
    Comparison of different implementations of Initial--Final-State 
    Interference. Numbers are relative to the cross-sections.}
 \label{fig1}
\end{figure}

\begin{figure}[htb]
  \begin{center}\mbox{
      \epsfig{file=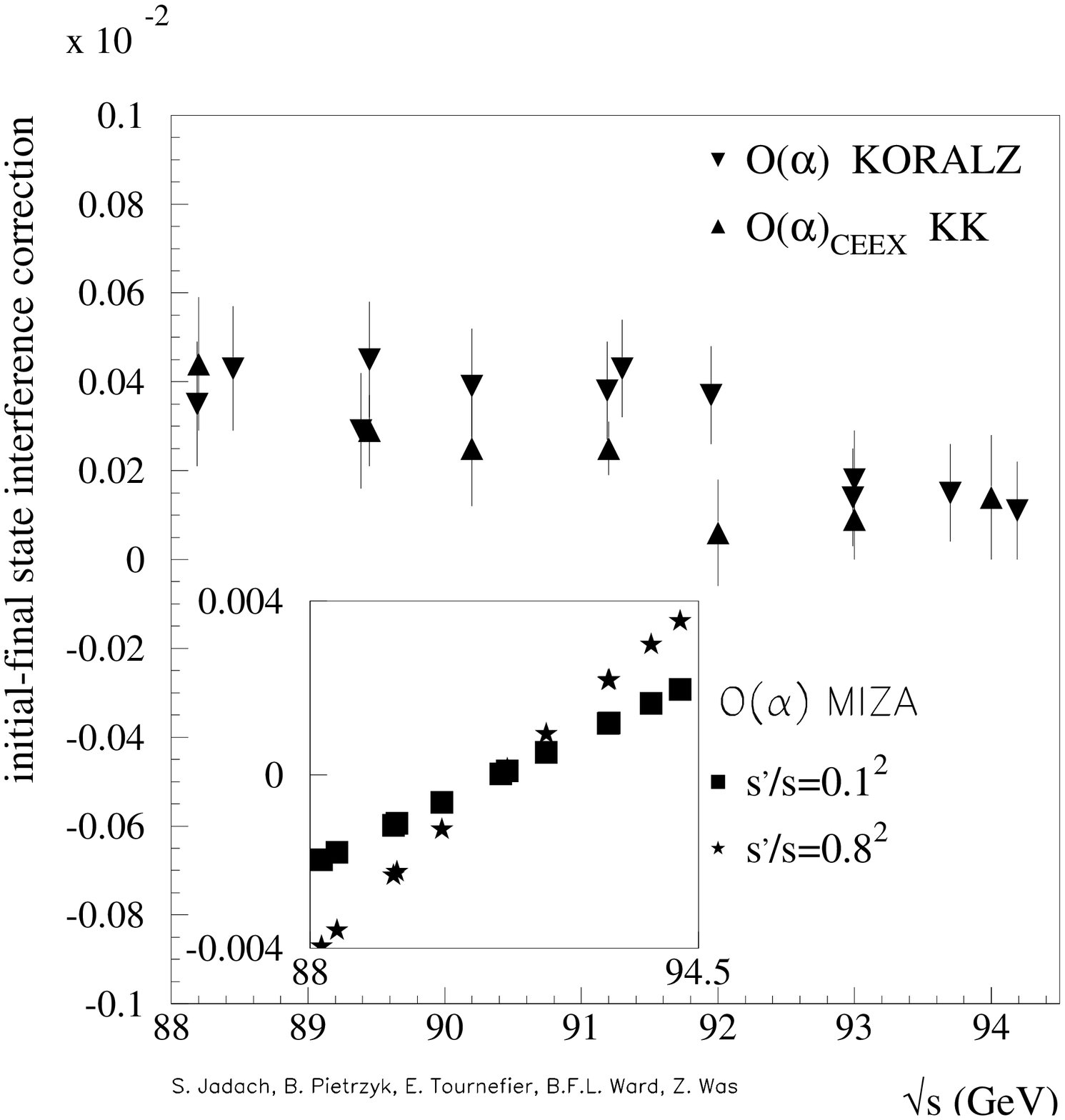,width=15cm}}
  \end{center}
  \caption[]{\sf
    Corrections to the ALEPH experimental efficiency due to implementation
    of the Initial--Final-State Interference (IFI) calculated with the \KK{} 
    and a
    modified version of the KORALZ MC programs. The inset shows the dependence
    of the IFI corrections on the $s$' cut. Note that both the vertical and 
    horizontal scales of the
    figure are different from those of the inset.
    }
  \label{fig2}
\end{figure}

It is important to stress
that the theoretical answer for pure \Order{\alpha^1} IFI is known {\em unambiguously}, 
since the exact formula for the $\gamma$-$Z$ box was published in Ref.~\cite{brown:1984}
and the exact single real bremsstrahlung matrix element was
given in Ref.~\cite{Berends:1981yz}.
Since then, the only problem left is purely technical, i.e. the phase-space integration.
Contrary to other calculations\footnote{
  The IFI calculation for an acollinearity cut-off
  in ZFITTER  is based on certain approximations, albeit justifiable near Z.}, 
the pure \Order{\alpha^1}  subgenerator of KORALZ%
\footnote{
  The pure \Order{\alpha^1}  subgenerator of KORALZ is based on
  the MC of Ref.~\cite{Mustraal}; 
  however, this was never tested below a 2\% precision level,
  and it contained the approximate $\gamma$-$Z$ box.}
features the {\em exact} phase-space integration.
Furthermore, in checking the exact analytical calculations
in Ref.~\cite{JKSW} for the IFI,  results from \Order{\alpha^1} KORALZ were cross-checked 
to the precision level of $10^{-4}$. In refs. \cite{Jadach:1991va,Was:1990ce}
(mainly, but not exclusively, devoted to forward--backward asymmetry), 
some important details on the technical side of these
high-precision tests are also given.
This \Order{\alpha^1}  subgenerator of KORALZ is at present
the best available benchmark for the  calculation of the \Order{\alpha^1} IFI,
for {\em any} experimental cuts.

Measured experimental cross-sections are extrapolated to full angular acceptance
and to low $s'$ before the Z line-shape fit is performed. The published
cross sections for the ALEPH Collaboration 
\cite{ALEPH} correspond to the $s'/s$ cut of (0.1)$^2$ for hadronic final states 
and to (2m$_\tau)^2/M_Z^2$ for leptonic final states. 
Figure~\ref{fig1} shows good agreement between different  ${\cal O}$($\alpha^1$) 
calculations of IFI for the leptonic Z decays 
in MIZA \cite{MIZA}, using formulas from Ref.~\cite{JKSW}, 
ZFITTER \cite{ZFITTER}, as described in \cite{PCP},
and, what is very important, also with the \Order{\alpha^1} of KORALZ \cite{KORALZ_4.0}.

\begin{table}[htb]
  \begin{center}
    \begin{tabular} {|l|c|} \hline
      Test option & Shift of $M_Z$  \\ \hline
      No IFI & $+$0.04 MeV  \\
      Only lepton IFI & $-$0.17 MeV \\
      Only hadron IFI & $+$0.17 MeV \\
      Lepton and 50\% hadron IFI & $-$0.07 MeV \\ \hline
    \end{tabular}\end{center}
  \caption{\sf
    Change of the $M_Z$ on the line-shape fit output with different
    implementations of the Initial--Final-State Interference (IFI)  correction
    with respect to the full calculations including IFI correction.}
  \label{shifts}
\end{table}

The IFI correction for hadronic Z decays is about ten times smaller, owing to 
partial cancellation of contributions from different quark flavours.
Relative IFI correction to both leptonic and hadronic  cross section
is very well approximated by a straight line, with zero at peak position.
It influences, therefore, only the fitted value of the Z mass 
$M_Z$ \cite{ManelBolek}. As seen in Table~\ref{shifts}
the absolute contributions of IFI to both hadronic and leptonic cross sections,
and therefore to the fitted $M_Z$ are simmilar.
The slope of the IFI correction in Fig.~\ref{fig1}
in the hadronic and leptonic cases is opposite and the combined 
effect of the respective contributions to $M_Z$ is negligible.

This conclusion is so far limited to \Order{\alpha^1} and the important question is:
Can higher orders change the IFI-integrated cross section significantly?
Let us investigate the basic properties of IFI before we go into details.
The important properties of  IFI corrections at and around the $Z$ peak, as seen 
explicitly 
in the \Order{\alpha^1} analytical results, are that 
(a) the IFI correction is suppressed by a factor $\Gamma_Z/M_Z$ 
(or $(s- M_Z^2) / M_Z^2$),
(b) for $e^-e^+\to f\bar{f} $ it does not contain mass logarithms 
    $\ln(s/m_e^2)$ or $\ln(s/m_f^2)$.
It is of order
\begin{displaymath}
  {\sigma^{IFI}\over \sigma} \sim   Q_eQ_f {\alpha\over\pi} 
{\rm Max} \Bigl\{ {\Gamma_Z\over M_Z}\; ; \;  {s- M_Z^2\over M_Z^2} \Bigr\}
\end{displaymath}
where $Q_e$ and $Q_f$ are electric charges of the electron and final-state fermion.
The above correction is therefore numerically small, provided
cuts on photon energies are absent or loose.
The reason for the ${\Gamma_Z/ M_Z}$ suppression is the time separation
between the production and decay processes of the relatively long-lived Z,
while the reason for the absence of mass logarithms is that the IFI interference
comes from the angular range where the  photon is at roughly equal angular distance from both
initial- and final-state fermions%
\footnote{ 
  This can be clearly seen
  for real photons, and the same has to be true for the virtual ones (IR cancellations).
}.
The above two reasons are very elementary and simple; they are not limited to
any perturbative order: they will lead to the same suppression pattern at higher orders.
If in the actual calculation they do not, 
then it is only because of some bad unphysical and/or technical approximation
made in the perturbative calculations, which will be cured
in a more complete calculation.
For instance, the overall size of the interference correction is the result
of delicate cancellation of a rather big infrared (IR) infinite contribution
from real and virtual QED corrections.
If these cancellations are disturbed, the higher orders may get artificially enhanced.
Typically, if the \Order{\alpha^1} IFI corrections, in an attempt at
 including  big non-interference higher orders,
 are improperly ``folded in''
with the big non-interference higher-order ones, then
we may get an unphysical enhancement of the IFI corrections.
Of course, we expect some interplay of the IFI with big non-interference corrections.
However, this should generally happen in a ``multiplicative way'', 
such that the real-virtual cancellations for the interference contributions
are maximally preserved.

One example of improper technical approximation enhancing IFI corrections
is already known within \Order{\alpha^1};
in the calculation of Ref.~\cite{Mustraal}, the authors have applied the so-called
``pole approximation'', generally accepted and known to work well, 
for the box diagram and soft bremsstrahlung. 
This damages the delicate cancellation between the two, 
resulting in a falsely increased interference correction around the $Z$ peak
(but not at the peak itself).
This approximation does not really violate the $\Gamma/M_Z$ suppression of IFI.
It only causes the coefficient in front of $(s- M_Z^2)/ M_Z^2$ to be incorrect.

What is the generic size of the IFI at \Order{\alpha^2} near the $Z$?
We expect the contribution to be of order:
\begin{equation}
\label{correct}
  {\sigma^{IFI}\over\sigma} \sim  
{\rm Max} \Bigl\{ {\Gamma_Z\over M_Z}\; ; \;  {s- M_Z^2\over M_Z^2} \Bigr\}
  Q_eQ_f \left({\alpha \over \pi}\right)
  \times
  Q_e^2 \left({\alpha \over \pi}\right)
  \ln{s \over m_e^2} \ln{\Gamma_Z\over E_{beam}},
\end{equation}
where the
$(\alpha / \pi) \ln(s/ m_e^2) \ln(\Gamma_Z/E_{beam})$ part
is the Sudakov double logarithm from the non-interference ISR,
essentially the same which reduces by 30\% the Z peak cross section at \Order{\alpha^1}.
The $\ln(\Gamma_Z/E_{beam})\sim \ln(E_{\max}/E_{beam})$ 
is due to the cut on ISR
photon energy induced by a resonance behaviour in the Born cross section.

In  Fig.~\ref{fig1} we show results for the IFI contribution to the
total muonic and hadronic cross section from the \Order{\alpha^1} calculations
of MIZA, ZFITTER, \Order{\alpha^1} KORALZ and from the 
\KK\ MC~\cite{KK:1999}, which is the only available calculation beyond  
\Order{\alpha^1} (see below for more details).
As we see, all four \Order{\alpha^1} results agree fairly well, as they should 
(only a technical problem could cause a difference)
while results from the exponentiated  {\em Coherent Exclusive Exponentiation} (CEEX) \Oceex{\alpha^1},
 calculation of \KK\ MC
differs from previous one in the slope of the energy dependence by about 25\%{}\footnote{
  Different conventions appear in the literature on 
how the {\it relative}
interference correction should be presented. In our case we calculate the
correction always within the given order of perturbation expansion. Thus, e.g.
  $\delta_{int}(\Orderd{\alpha^1})=
     [\sigma_{int}(\Orderd{\alpha^1}) - \sigma_{no\; int}(\Orderd{\alpha^1}) ]/
     \sigma_{no\; int}(\Orderd{\alpha^1}) $
  and
  $\delta_{int}(\Oceexd{\alpha^1})=
  [\sigma_{int}(\Oceexd{\alpha^1}) - 
    \sigma_{no\; int}(\Oceexd{\alpha^1}) ]/ 
    \sigma_{no\; int}(\Oceexd{\alpha^1}) $.
  }.
This is exactly the size we expect from eq.~(\ref{correct}).
Note that for leptons and  \Oceex{\alpha^1} the IFI correction has its zero
at a higher value of $\sqrt{s}$ than  for \Order{\alpha^1}. 
For hadrons the above effect is less pronounced, because of
cancellations between different flavours.

Let us now briefly explain why exponentiation is the correct and economical solution
for getting a hint on IFI beyond \Order{\alpha^1}.
Already from the \Order{\alpha^1} we learn one important lesson:
the IFI corrections are large if we apply a strong cut on photon energies.
This is the only possible big enhancement factor at this order.
In fact these corrections may be so large that it is necessary to sum them
up to infinite order, i.e. to exponentiate.
This is precisely what is done in the \KK\ MC, 
with the {\em Coherent Exclusive Exponentiation} (CEEX)
described in Ref.~\cite{Jadach:1998jb}, employing spin amplitudes of Ref.~\cite{GPS}%
\footnote{
  In the first multiphoton MC with exponentiation beyond \Order{\alpha^1}
  of Ref.~\cite{yfs2:1990}, which is used in the multiphoton KORALZ, 
  the IFI is completely omitted.}.
Why is exponentiation the correct approach to IFI beyond \Order{\alpha^1}?
First of all, because exponentiation is basically ``multiplicative''
and it will thus preserve at higher orders the virtual--real cancellations that are at the heart
of the IFI suppression.
Secondly, it sums up the only possible enhancement factor in IFI, which is due to 
the strong cut ($\sim\Gamma_Z$) on photon energy.

In practice one has to be careful.
For instance, also the important virtual correction \cite{greco:1980}
$Q_eQ_f (\alpha/\pi) \ln(t/u) \ln((M_Z-iM_Z\Gamma_Z-s)/M_Z)$
has to be summed up to infinite order (included in the exponential form factor).
As discussed in Ref.~\cite{Jadach:1998jb}, failure to do this
leads to a disturbance of the virtual--real cancellation and consequently
to lack of the suppression factor $(\Gamma_Z / M_Z)$ at \Order{\alpha^2},
resulting in a dramatic unphysical increase of the IFI correction close to 
the Z peak.

Let us finally note that also the \KK\ MC~\cite{KK:1999} provides 
interesting results on IFI for LEP2 energies. They will be
presented in a separate publication \cite{ifiLep2:1999}.

Another source of the theoretical systematics is related to QCD corrections for 
the hadronic Z decays, as well as vacuum polarization.
The above discussion of IFI is made  in the framework of pure QED. 
However, quarks may emit final-state gluons as they emit 
initial- and/or final-state photons. 
The correction of the lowest possible order related to the above configuration,
taking only the QED IFI part, is expected to be 
\begin{displaymath}
   {\delta\sigma^{IFI}\over\sigma} \sim  
   {\Gamma_Z\over M_Z} Q_eQ_q{\alpha\over \pi} \times
   {\alpha_s \over \pi} \sim 4\times 10^{-6}
\end{displaymath}
provided we do not apply very sharp acceptance cuts 
on photon and gluon energies
(or do not try to look into a sub-sample with isolated photon or hard gluon).
This correction cannot be enhanced by the logarithms of the electron 
mass $|\ln(m_e^2/s)| \sim 20$.
This is because
the IFI contribution can only arise from the phase-space region
where the photon angle from initial- and final-state fermions is large and about the same, 
while the mass logarithms $\ln(m_e^2/s)$ and $\ln(m_f^2/s)$
come mostly from the small photon angles.
This mechanism cannot be changed by hadronization of additional gluon emission because
hadrons in jets are narrowly correlated and a jet is acting coherently 
as an  effective object of almost zero or low electric charge.
None of the perturbative or non-perturbative self-interaction within a jet 
can change this simple fact. Nor can the
gluon emission induce a logarithm of the quark mass,
even in the presence of an additional photon, because of the Kinoshita-Lee-Nauenberg theorem.

We shall assume here that these QED--QCD corrections can change
the \Order{\alpha^1} IFI correction for hadronic final states
by the rather conservatively large amount of 50\%.
The Z mass on the MIZA fit output is then changing by less than 0.1 MeV, 
when the hadronic IFI  correction is changed by 50\% as seen in Table~\ref{shifts}.
In this way the theoretical error due to the possible 
\Order{\alpha_s \alpha_{QED}} correction largely
dominates the other possible theoretical uncertainties discussed above. 
Therefore the error of 0.1 MeV on the $M_Z$ is the total conservative theoretical uncertainty 
on the IFI calculations.

\begin{table}[htb]
  \begin{center}
    \begin{tabular} {|l|c|c|c|} \hline
            & $\sigma^0_{had}$  & $M_Z$    & $\Gamma_Z$  \\ \hline
      ISR   & $1\times10^{-4}$   & Negl.    & Negl.       \\
      Pairs & $1.8\times10^{-4}$ & 0.1 MeV  & 0.1 MeV     \\
      IFI   & negl. & 0.1 MeV   & negl.                  \\  \hline
      total & $2\times10^{-4}$   & 0.15 MeV  & 0.1 MeV    \\ \hline
    \end{tabular}\end{center}
  \caption{\sf
    Theoretical uncertainties in the MIZA line-shape fit results coming
    from the precision of the calculation of Initial State Radiation ISR 
    \cite{StBoMa}, fermion pair production \cite{StBoMa}
    and Initial--Final-State Interference IFI.}
  \label{uncertainties}
\end{table}
Theoretical uncertainties on the MIZA line-shape fit results coming
from the precision of the calculation of ISR \cite{StBoMa},
fermion-pair production \cite{StBoMa} 
and IFI are given in  
Table~\ref{uncertainties}.

The precision of the IFI contribution is discussed above, 
including Fig.~\ref{fig1} for the  low values of the $s'/s$ cut
used for the published values of the cross-section measured by LEP. 
These low values are similar to the ones used for the hadronic event selection by the LEP
experiments, but they are much lower than the ones used typically for the leptonic selection.  
A lepton acollinearity cut of 20$^\circ$ is made by the ALEPH Collaboration
\cite{ALEPH}, corresponding approximately to $\sqrt{ s'} > 0.8 \sqrt s$. 
Calculation of the experimental efficiency 
using  a program without implementation of the
IFI contribution could lead to an important experimental bias, as seen in 
the inset in Fig. \ref{fig2} where ${\cal O}$($\alpha^1$) calculations 
in MIZA \cite{MIZA} (using formulas from
Ref.~\cite{JKSW}) are compared for two different values of the $s'/s$ cut. 
The corrections for the experimental efficiency could be as high as the 
difference between these two distributions in the case of a full geometrical
acceptance of the detector. The real corrections are, however, smaller, owing
to angular cuts. For example, in the ALEPH leptonic selection it is 
required that $|\cos\Theta^*|<0.9$, where $\Theta^*$ is the centre of mass 
scattering angle. 
The IFI is definitely bigger at high $|\cos\Theta^*|$.
Figure \ref{fig2} shows the corrections to the ALEPH muon  efficiency
calculated using the \KK{}  MC program. 
This correction is equal to the relative difference of the acceptance 
calculated with and 
without IFI.
The results of the 
${\cal O}$($\alpha^1$) calculations using a modified  version of the KORALZ MC 
program are similar. The necessary modifications, which are  absent in the public 
version of the code, consist of the use of weighted events allowing 
the shift of the so-called
$k_0$ cut-off to an arbitrary low value.
This enables one to remove the related bias; 
see refs. \cite{Was:1990ce,Jadach:1991va}
and references therein for technical details. The $k_0$ cut-off 
separates the region of the phase space 
where the kinematics of the 
event includes the hard photon 4-momentum from the one where it is integrated
and summed with virtual corrections. 

\vspace{5mm}
\centerline{\bf\large Acknowledgements}
\vspace{2mm}
Part of this work was carried out in the framework of the collaboration between 
IN2P3 and the Polish laboratories and supported in part by Polish Government 
grants KBN 2P03B08414 and KBN 2P03B14715 and the Maria Sk\l{}odowska-Curie 
Joint Fund II PAA/DOE-97-316. This work was also supported in part 
by the US DoE contracts DE-FG05-91ER40627 and DE-AS03-76SF00515. 
S.J., B.F.L.W. and Z.W. thank the CERN
TH and EP Divisions and all four LEP Collaborations for their support.

\end{document}